\documentclass[aps,pre,preprint,showpacs,groupedaddress]{revtex4}
\usepackage[dvips]{graphicx}
\begin{document}

\title{Effect of edge removal on topological and functional robustness of complex networks}

\author{Shan He}
\author{Sheng Li}
\email{lisheng@sjtu.edu.cn}
\author{Hongru Ma}
\affiliation{Department Of Physics, Shanghai Jiao Tong University, Shanghai 200240, People's Republic of China}

\date{\today}

\begin{abstract}
We study the robustness of complex networks subject to edge removal. Several network models and removing strategies are simulated. Rather than the existence of the giant component, we use total connectedness as the criterion of breakdown. The network topologies are introduced a simple traffic dynamics and the total connectedness is interpreted not only in the sense of topology but also in the sense of function. We define the topological robustness and the functional robustness, investigate their combined effect and compare their relative importance to each other. The results of our study provide an alternative view of the overall robustness and highlight efficient ways to improve the robustness of the network models.
\end{abstract}


\maketitle

\section{introduction}
Complex networks are ubiquitous in our world. They exhibit not only diverse structural characteristics~\cite{ref1,ref24}, such as the power-law tail of degree distribution and the small-world phenomenon of average path length, but also different levels of robustness, e.g., scale-free networks display higher tolerance to error but more vulnerability to attack than exponential networks~\cite{ref5,ref6,ref7,ref16}. Robustness evaluates the ability of a network to maintain its original attributes and functions when constituent loss or other kinds of damage are present. Once the robustness is identified, the weakness of a network is pointed out for optimizations or countermeasures, e.g., the weakness of a communication network can be overcome to increase reliability, while the weakness of an epidemic network can be utilized for efficient destruction. For these practical applications, the study of the robustness have received a lot of interests.

Several measures of the robustness have been proposed. A frequently used one is the existence of the giant component~\cite{ref5,ref6,ref7,ref8,ref16}. The giant component is the only component in a network whose size scales linearly with the number of vertices. It was found that the damage to a network, such as random removal of vertices or edges, can be exactly mapped to a standard percolation process. The network percolates if the giant component exists, indicating that the general connectedness of the network is maintained. There are also other measures based on quantities such as efficiency~\cite{ref8,ref15} and average path length~\cite{ref5,ref8}. The measure we adopt in the present work is the preservation of total connectedness~\cite{ref9}. The total connectedness of a network is preserved only when communication is effective between every pair of vertices in the network. The more damage needed to destroy the total connectedness, the more robust the network is. This measure can be useful to describe the robustness of networks that have no vertex redundancy. In such networks, each vertex contributes to the whole in a way that cannot be replaced by the others and even the unavailability of a single vertex affects the overall functionality, e.g., in a scenario that many computers collaborate through a network to accomplish some calculation-intensive task, if one or more members lose communication, the performance of the collaboration may be degraded or the task may even fail. Note that the total connectedness used in the measure is not limited within the sense of topology, i.e., physical connection is not the only factor that influences the effectiveness of the communication; there are other factors that can hinder the communication even though all the vertices are connected. These factors are often related to the function of a network, e.g., it was reported that in a transmission network, a path connecting two vertices may become so long after damage that this path is unusable~\cite{ref17}. As both topology and function related factors can prevent the communication between vertices, two types of robustness are involved: the topological robustness and the functional robustness. Studying their combined effect and the dominance of one over the other can provide additional insights into the network robustness. However, such studies have not yet been carried out extensively.

Besides the measures, the damage of constituent loss has also been widely studied in the form of vertex removal. Removing a vertex is an appropriate abstraction of several real-world events, e.g., a user leaves a P2P network, a website goes offline. There are other events that should be abstracted more appropriately by removing an edge, e.g., a network cable is unplugged, a flight between two cities is canceled. However, only a few works cover edge removal. In fact, there are remarkable differences between the two types of removal. In the sense of topology, vertex removal inflicts more damage, as each time a vertex is removed, all of its edges are removed as well. Thus to achieve the same effect, smaller amount of vertices is removed than edges. If the damage of removal is assessed in terms of edges rather than vertices, the vulnerability of a network could be moderated~\cite{ref13}. In the sense of function, the removal of vertices reduces the total amount of quantity transmitting on a network, as vertices usually not only deliver but also generate the quantity. If the reduction of the total amount can compensate the damage, the robustness of the network is enhanced~\cite{ref9}. On the other hand, the enhancement may not be expected for edge removal because there is no change in the total amount. These differences indicate that edge removal needs separate studies from vertex removal.

In this paper, we propose to study the effect of edge removal on the topological robustness and the functional robustness, using the total connectedness as the criterion of breakdown. Several network topologies and removing strategies are simulated. We combine a simple traffic dynamics with the network topologies and successively remove edges until the total connectedness of the networks is destroyed. The fractions of removed edges characterize the robustness of the networks. Moreover, as the destroy of the total connectedness is related to either the topological factors or the functional factors, we see which type of the factors is dominant. The purpose of our study is to explore the relationship between the topological robustness and the functional robustness, which is different from the purposes of previous works that also deal with edge removal~\cite{ref11,ref19,ref20,ref21,ref30}. Particularly, we only concern the times when the networks lose the total connectedness, and do not discuss the evolutions thereafter, such as the occurrence of cascading failure~\cite{ref14}. The results of our study provide an alternative view of the network robustness and highlight efficient ways for improvements.

The paper is organized as follows: In Sec. II, the model we studied is described in detail. In Sec. III and IV, the simulation results are shown according to the removing strategies. In Sec. V, we give the conclusion.

\section{network topologies, robustness and removing strategies}
The network models of topology we studied are the Erd\"os-R\'enyi (ER) model of random network~\cite{ref1}, the Barab\'asi-Albert (BA) model of scale-free network~\cite{ref2} and the Newman-Watts (NW) model~\cite{ref4}, which is a variant of the Watts-Strogatz (WS) model of small-world network~\cite{ref3}. For all the network models, we set vertex number $N=1024$ and average degree $\langle k\rangle =8$. The edges are undirected and have no weight. Multiple edges are not allowed. In order to study the robustness, we ensure that the topologies are totally connected when they are intact. The ER model yields exponential degree distribution. Each vertex pair is linked by an edge with probability $\langle k\rangle /(N-1)$. Since $\langle k\rangle >\ln (N)$, the ER networks are almost surely connected~\cite{ref1}. The BA model features growth and preferential attachment. At any time step of a growth, the network from the previous step is connected. A new vertex is added with $\langle k\rangle /2$ preferentially linked edges in the current step and the connectedness is preserved. If a BA network grows infinitely large, the probability of a randomly selected vertex having degree $k$ is proportional to $k^{-3}$. The NW model starts with a regular structure, which is identical to the WS model. But unlike the WS model, the NW model builds shortcuts by randomly inserting $pN$ edges, where $p$ is the rate of the shortcuts. This way of building shortcuts eliminates the possible network fragment during the process of rewiring in the WS model~\cite{ref4}. As total connectedness is the special concern, we adopted the NW model instead of the WS model. With $p$ changes from small to large, a NW network can undergo a transition from ``large world'' to ``small world''. We choose $p=0.02$ in the small-world phase without loss of generality. The $pN$ shortcuts are inserted into a periodic square lattice with side length $L=32$ and coordination number $z=8$. As $p$ is a small quantity, the average degree is $8+p\approx 8$. Among the three models, the BA model produces scale-free degree distribution which is observed in many real networks~\cite{ref1}. The NW model is based on a square lattice, which is the topology that can reproduce some of the observed real Internet features~\cite{ref31}. The randomly inserted shortcuts are for the small-world property, though high clustering is absent. The ER model does not match real networks in nearly all aspects. Nevertheless, it has the significance that many of its properties can be obtained through probabilistic approaches. Note that the average path lengths of all the three models scale logarithmically with the number of vertices. It is interesting to compare the robustness of them.

The robustness of the network models refers to the ability of maintaining total connectedness when a portion of edges are removed. The total connectedness is defined as the effective communication between every pair of vertices, which is affected by the twofold effects of the edge removal. One is that the physical connection between a pair of vertices may be cut off. If the number of edges in a network is decimated to be less than $N-1$, the network is certainly not totally connected. The other effect is related to the fact that a network usually performs some function, e.g., transmits some quantity. The edge removal may redistribute the quantity transmitting on the network and the vertices may lose communication due to congestion~\cite{ref10}, e.g., the quantity from a source vertex is congested on some intermediate vertices and never reaches the target. The former effect is topological, while the latter is functional. Either of the two effects can destroy the total connectedness, and cause the breakdown of the networks.

The topological breakdown of the networks occurs when the networks are just fragmented from one component into two during the edge removal, i.e., at least one of the vertices is isolated physically from the others~\cite{ref9}. This condition is more restrictive on the connectivity than the existence of the giant component, which was usually used in previous works~\cite{ref5,ref6,ref7,ref8,ref13,ref14,ref16}. We adopt this condition in favor of the emphasis on the completeness of a network and that the functional breakdown can be defined in a similar manner.

We incorporate a simple traffic dynamics to define the functional breakdown of the networks. At each time step, a certain amount $\rho $ of quantity, which can be data, energy, etc., is transmitted between every pair of vertices, along the shortest paths. Transmitting along the shortest paths minimizes the delay of distributing the quantity. For the logarithmic average path length of the network models, the transmission scales well. The number of shortest paths passing through a vertex $j$, which defines the vertex betweenness $B_j$ of $j$~\cite{ref26}, determines the load $L_j$ of the transmitting quantity on $j$, $L_j=\rho B_j$~\cite{ref10}. The load is handled by the vertex within its capacity, i.e., the amount of quantity that can pass through $j$ is at most the capacity of the vertex. It would be efficient to set the capacity of $j$ proportional to $B_j$, so as to meet the load. However, the calculation of betweenness needs global information and takes time $O\left(\langle k\rangle N^2\right)$~\cite{ref22}, which is resource consuming. For simplicity, we set uniform capacity $C$ for each vertex. As a result, the vertex with the largest betweenness $B_{\max }$ is the most likely to encounter overload. The largest generation rate $\rho_c$ of the transmitting quantity is $C/B_{\max }$. The rate $\rho_c$ marks the capacities of the networks. It was found that homogeneous networks have higher capacities than heterogeneous ones~\cite{ref10}. As the purpose of the study is to compare the different aspects of the robustness rather than the absolute capacities, we normalize $\rho$ with respect to $\rho_c$ for each individual network, i.e., the load of $j$ changes to $L_j=\rho B_j C/B_{\max }$, where $\rho$ is the normalized generation rate of the quantity. The load $L_j$ is initially less than $C$ with $0<\rho <1$, but $L_j$ could be larger than $C$ as the vertex betweenness redistributes in the process of edge removal. If removing an edge leads to $L_j>C$ for some vertex $j$, congestion occurs~\cite{ref10}. The excessive load accumulates continuously on $j$ and the vertex is unable to communicate with other vertices. Similar to the topological breakdown, the total connectedness is destroyed. This situation defines the functional breakdown of the networks. One can see that the functional breakdown actually depends on the evolution of the ratio $B_j/B_{\max }$; the study of the functional robustness is essentially a test of the response of vertex betweenness to edge perturbations. Though the dynamics is simple, some essential functional properties of the networks are reflected and we are able to study the relative importance of the functional robustness to the topological robustness under different values of the only parameter $\rho $.

To study the two types of the robustness, we start from an undamaged network generated by one of the network models and remove the edges one at a time until the network breaks down for either the topological or the functional reasons. The robustness are studied in the contexts of different removing strategies.

Two removing strategies are used: random failure and attack. The random failure strategy removes edges with uniform probability, which can be seen as a simple abstraction of the successive error in a communication network. The attack strategy removes edges in the descending order of their importance, which tries to model a sophisticated attacker who knows the global information of a network and always targets the most important link. There are several quantities that can define the importance of an edge, such as edge betweenness and edge degree~\cite{ref30,ref8}. We choose edge betweenness as the definition of the importance, because betweenness directly measures the load of the transmitting quantity in our traffic dynamics. Analogous to vertex betweenness, edge betweenness is defined as the number of shortest paths passing through an edge~\cite{ref30}. Every time the most important edge is removed, we recalculate the betweenness to find the most important one in the remaining edges. If there are multiple edges having the same largest betweenness, we randomly select one for removal. Random failure and attack are usually studied together to obtain a collective characterization of network robustness~\cite{ref5}. In the next two sections, we show the results according to the two strategies.

\section{the random failure strategy}
\begin{figure}
\centering
\includegraphics[width=86mm]{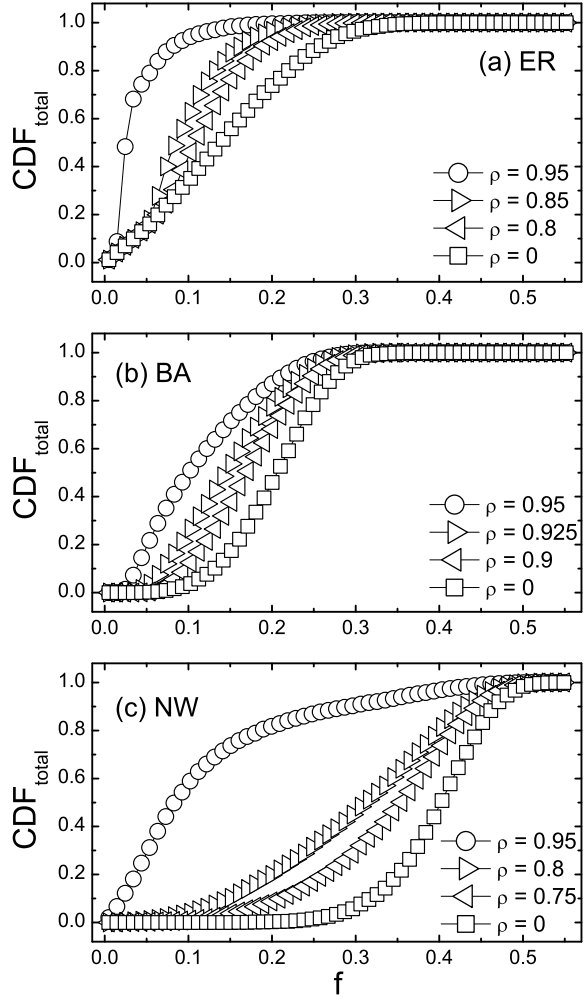}
\caption{The cumulative distribution function $CDF_{total}$ of the total breakdown under the random failure strategy as a function of the fraction $f$ of removed edges for (a) the ER model, (b) the BA model and (c) the NW model. When $\rho = 0$, the total breakdown is in fact the topological breakdown; when $\rho >0$, the total breakdown is the combined effect of the topological breakdown and the functional breakdown. All the data is averaged over $10240$ realizations.} \label{figure1}
\end{figure}

In this section, edges are randomly removed from the networks. We first study the total breakdown, which is the combined effect of the topological breakdown and the functional breakdown. Figure \ref{figure1} shows the cumulative distribution function $CDF_{total}$ of the total breakdown as a function of the fraction $f$ of removed edges for the network models. The results are shown for different values of $\rho$.

When $\rho =0$, there is no quantity transmitting on the networks. The total breakdown is in fact the topological breakdown, which is characterized by a critical fraction $f_c^{topo}$ of removed edges. When $f<f_c^{topo}$, each of the networks is totally connected; when $f\geq f_c^{topo}$, each of the networks is split into two or more components, where one of the components could be the giant component.

For the ER model, almost every model realization is connected if $\langle k\rangle >\ln (N)$~\cite{ref1}. Thus $f_c^{topo}=\left[\langle k\rangle -\ln (N)\right]/\langle k\rangle $.

For the BA model, a simple estimate of $f_c$ is sought under the framework of random graph with prescribed degree sequence~\cite{ref29}. Seeing that the BA model yields no assortative mixing~\cite{ref28}, i.e., no degree correlation, we equate the BA networks to random scale-free networks. Although the BA model features network assembly and evolution instead of complete randomness, some qualitative result can be obtained by the estimate and confirmed through simulation. For a random network with arbitrary degree distribution $P(k)$, the probability $\pi _s$ of a randomly selected vertex being in a finite component of size $s$ has been obtained in Ref.~\cite{ref27},
\begin{equation}
\pi _s=\frac{\langle k\rangle }{(s-1)!}\left[\frac{d^{s-2}}{dz^{s-2}}\left[g_1(z)\right]^s\right]_{z=0},
\end{equation}
where $g_1(z)=\sum _{k=0}^{\infty } \frac{(k+1)P(k+1)}{\langle k\rangle }z^k$ and $s>1$. Particularly, $\pi _1=P(0)$. We now consider a scale-free random network, which has the same $N$, $\langle k\rangle$ and $P(k)$ as the BA networks. With the minimum degree larger than $2$, $g_1(z)$ has no constant terms and $\pi _s=0$. There are no finite components, all the vertices are connected in the giant component. It was reported that the giant component always exists for a random network with $P(k)\sim k^{-3}$ when vertices are randomly removed~\cite{ref6}. The result is similar if edges are removed~\cite{ref13}. Thus the network is always split into the giant component and a finite component. Denote $\pi _s(f)$ the distribution of the size of the finite component after a fraction $f$ of edges are removed. When $f\to 0$, the network is nearly not affected and $\pi _s(f)\to 0$. Though $\pi _s(f)$ is small, various sizes of the finite components are probable. With the existence of the giant component, $\pi _s(f)\approx\pi _s(0)$ decays exponentially~\cite{ref27}. We then neglect the higher order components and focus only on the components of size one and two. The new degree distribution and the new average degree after edge removal are $P_f(k)=\sum _{k_0=k}^{\infty } P\left(k_0\right){k_0 \choose k}(1-f)^kf^{k_0-k}$ and $\langle k\rangle _f=(1-f)\langle k\rangle $, respectively~\cite{ref6}. We calculate the ratio
{\setlength\arraycolsep{2pt}
\begin{eqnarray}
r(f) & = & \frac{\pi _2(f)}{\pi _1(f)}=\frac{\left[P_f(1)\right]^2}{\langle k\rangle _fP_f(0)} \nonumber\\
& = & \frac{(1-f)\left[\sum _{k=1}^{\infty } k P(k)f^k\right]^2}{f^2\langle k\rangle \sum _{k=1}^{\infty } P(k)f^k}.
\end{eqnarray}
}This ratio is a monotonically decreasing function of $f$, which implies that in addition to the exponential decay of $\pi _s(f)$ when $f\to 0$, the components of size two become even less probable than the components of size one as $f$ becomes larger; in most cases, the network is split into the giant component and a finite component consisting of an isolated vertex. In the simulation of the BA model, we observed that in more than 99.8\% realizations, the finite component is of size one. This fact allows us to estimate the critical fraction $f_c^{topo}$ by the emergence of an isolated vertex, which has empty degree. If one such vertex can be sampled, the total connectedness is destroyed, i.e., $f_c^{topo}$ satisfies $P_{f_c{}^{topo}}(0)=\frac{1}{N}$.

\begin{figure}
\centering
\includegraphics[width=86mm]{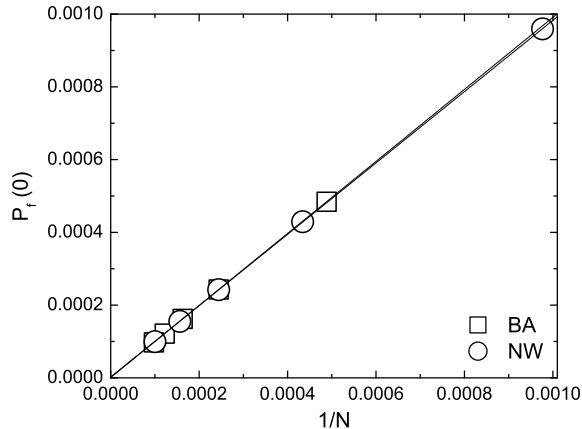}
\caption{The scaling relation between $P_f(0)$ and $1/N$ at $f=f_c^{topo}$ for the BA model and the NW model when edges are randomly removed. Symbols are the results of simulation, each of which is an average over $102400$ networks. The two solid lines (They nearly overlap.) are the linear fits of the simulation results. The slopes of the lines are $0.991$ and $0.980$.} \label{figure2}
\end{figure}

For the NW model, the critical fraction $f_c^{topo}$ is obtained in a similar manner. Employing the expression of $\pi _s(f)$~\cite{ref27}, we get
{\setlength\arraycolsep{2pt}
\begin{eqnarray}
\pi _s(f) & = & \frac{(1-f)^{s-1}\langle k\rangle }{(s-1)!}\left[\frac{d^{s-2}}{dz^{s-2}}\left[z^7\right]^s\right]_{z=f} \nonumber\\
& = & \frac{(7s)!\langle k\rangle f^{6s+2}(1-f)^{s-1}}{(s-1)!(6s+2)!},
\end{eqnarray}
}where the shortcuts in the model are neglected for convenience and the original degree distribution reads $P(k_0)=\delta (k_0-8)$. When $s$ is large,
\begin{equation}
\frac{\pi _s(f)}{\pi _{s-1}(f)}\approx \frac{7^7}{6^6}(1-f)f^6.
\end{equation}
This ratio tends to zero when $f\to 0$ and increases monotonically until it reaches the maximum value $1$ at $f=f_c=\frac{6}{7}$, where $f_c$ is the critical percolation fraction~\cite{ref6}. As the total connectedness is more restrictive than the existence of the giant component, we claim that $0<f_c^{topo}<f_c$. In this region, $\pi _s(f_c^{topo})$ drops exponentially, and again, each of the NW networks is split into the giant component and an isolated vertex, $f_c^{topo}$ satisfies $P_{f_c{}^{\text{topo}}}(0)=\frac{1}{N}$. Particularly, we get $f_c^{topo}\sim N^{-\frac{1}{8}}$ for the NW model. The result is consistent that if $N\to \infty$, then $f_c^{topo}\to 0$ and $\pi _s(f_c^{topo})$ decays very quickly with increasing $s$. In Fig.~\ref{figure2}, we plotted the scaling relation between $P_{f_c^{topo}}(0)$ and $1/N$ obtained in the simulation for the BA model and the NW model. As the figure shows, the agreement between the theoretical estimate and the simulation is reasonable. The above calculations show that $f_c^{topo}$ is strongly correlated to the local connectivity of vertices. In the BA model, the majority of the vertices have degree four, while in the NW model, all the vertices have degree at least eight. Thus as Fig.~\ref{figure1} shows, the NW model is more topologically robust than the BA model and they both are more topologically robust than the ER model.

\begin{figure}
\centering
\includegraphics[width=86mm]{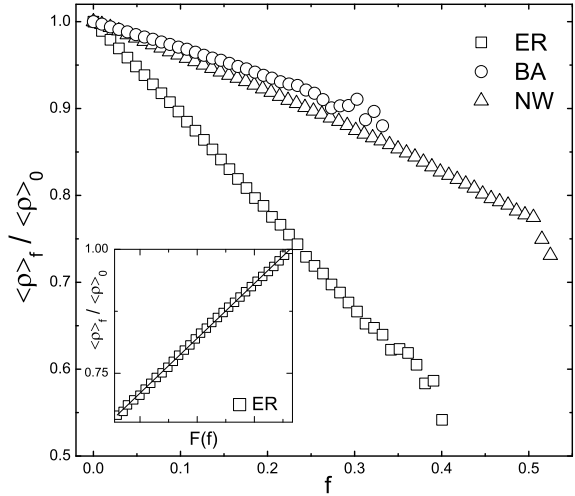}
\caption{The normalized average capacity $\left\langle \rho _c\right\rangle _f / \left\langle \rho _c\right\rangle _0$ under the random failure strategy as a function of the fraction $f$ of removed edges for the network models. The inset shows the linear relation between $\left\langle \rho _c\right\rangle _f / \left\langle \rho _c\right\rangle _0$ and $F(f)$ for the ER model, where $F(f)=k_{\max}(f)\ln[\left\langle k\right\rangle (1-f)]$. All the data is averaged over $10240$ realizations. The line in the inset is the linear fit.} \label{figure3}
\end{figure}

When $\rho >0$, the functional breakdown can also happen. Intuitively, networks with larger $\rho$ are more functionally vulnerable. We showed in Fig.~\ref{figure1} the cumulative distribution function of the total breakdown only for near capacity generation rates. When $\rho =0.95$, the ER networks are very likely to break down when only a small fraction of edges are removed. For the same $\rho$, the NW model has slightly larger $CDF_{total}$ than the BA model, which suggests that while the NW model is more topologically robust than the BA model, it is less functionally robust. Note in the figure that while $CDF_{total}$ for large $\rho$ increases fast in the region $0<f<0.1$ for all the network models, $CDF_{total}$ for $\rho =0$ increases much slower for the ER model or remains nearly zero for the BA model and the NW model. Thus in this region of $f$, the total breakdown is mainly determined by the functional breakdown. With $\rho$ close to $1$, the functional breakdown reflects the change in the network capacity during the edge removal. In Fig.~\ref{figure3}, we plotted the average network capacity $\langle \rho _c\rangle _f$ as a function of $f$ for the network models. ($\langle \rho _c\rangle _f$ is normalized to the initial value $\langle \rho _c\rangle _0$.) The capacity of all the network models decreases with the edge removal, because the shortest paths concentrate on the remaining edges and the betweenness of vertices becomes more heterogeneous.

For the ER networks, the degree distribution remains exponential when edges are randomly removed, i.e., the ER networks are still ER networks after the edge removal. The vertex betweenness $B$ of ER networks has a power-law relation $B\sim k^{\alpha}$ with degree $k$~\cite{ref23}, hence
{\setlength\arraycolsep{2pt}
\begin{eqnarray}
\left\langle \rho _c\right\rangle _f & = & \frac{C}{B_{\max}(f)}=\frac{C}{\sum _jB_j(f)\frac{\left[k_{\max}(f)\right]^{\alpha }}{\sum _k k^{\alpha }}} \nonumber\\
& = & \frac{C}{\sum _j B_j(f)\frac{(1+\alpha )\left[k_{\max}(f)\right]{}^{\alpha }}{\left[k_{\max }(f)\right]{}^{\alpha +1}-1}} \nonumber\\
& \approx & \frac{C k_{\max }(f)}{\sum _j B_j(f)(1+\alpha )},
\end{eqnarray}
}where $B_{\max}(f)$ is the maximum vertex betweenness and $k_{\max}(f)$ is the maximum degree when a fraction $f$ of edges are removed. The vertex betweenness also satisfies $\sum _jB_j(f)=D N(N-1)$, where $D\sim\ln(N)/\ln(\langle k\rangle_f)$ is the average path length~\cite{ref1}. Then we get
\begin{equation}
\left\langle \rho _c\right\rangle _f\sim \frac{C k_{\max }(f)\ln \left(\langle k\rangle _f\right)}{(1+\alpha )N(N-1)\ln (N)}
\end{equation}
and the normalized average capacity
\begin{equation}
\frac{\left\langle \rho _c\right\rangle _f}{\left\langle \rho _c\right\rangle _0}\sim k_{\max }(f)\ln [\langle k\rangle (1-f)]=F(f),
\end{equation}
where $k_{\max}(f)$ and $f$ have an implicit relation~\cite{ref24} through the regularized gamma function $P(\alpha,x)$ that $P\left[\left\lfloor k_{\max }(f)+1\right\rfloor ,(1-f)\langle k\rangle \right]=\frac{1}{N}$. The linear relation between $\left\langle \rho _c\right\rangle _f / \left\langle \rho _c\right\rangle _0$ and $F(f)$ is also shown in Fig.~\ref{figure3}. The average network capacity of the ER model decreases much more quickly than those of the other two network models, thus the ER model has the worst functional robustness.

For the BA model, though there is a similar power-law relation $B\sim k^\eta$ in scale-free networks~\cite{ref25}, scale-free degree distribution does not remain scale-free when edges are randomly removed~\cite{ref20}. Instead, we compare the trend of the the load distribution for the BA model and the NW model. The two network models both have heterogeneous load distributions. Hub vertices or shortcuts carry large load. The BA model also has heterogeneous connectivity while the NW model has homogeneous connectivity. When each edge is removed with equal probability, the hub vertices in the BA networks are more likely to be diminished than the non-hub vertices, while the shortcuts in the NW model are not biased. The BA networks tend to have more homogeneous load distribution than the NW networks. As shown in Fig.~\ref{figure3}, the average capacity of the BA networks decreases slightly slower than that of the NW networks. The BA networks are a little more functionally robust.

\begin{figure}
\centering
\includegraphics[width=86mm]{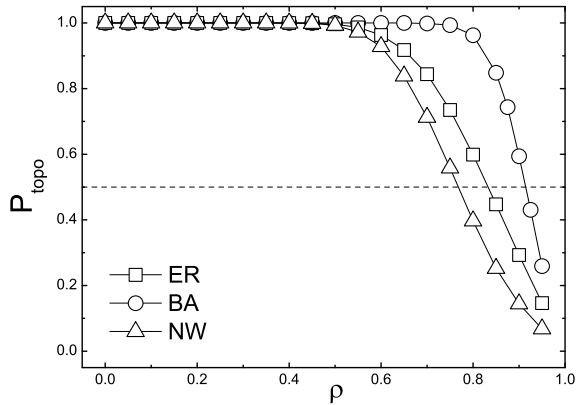}
\caption{The occurrence probability $P_{topo}$ of the topological breakdown under the random failure strategy as a function of the normalized generation rate $\rho$ for the network models. The data is averaged over $10240$ realizations. The dashed line is for $P_{topo}=50\%$. The network model that corresponds to the smallest $\rho$ at the point of intersection between $P_{topo}$ and this line has the best relative topological robustness.} \label{figure4}
\end{figure}

After studying the combined effect, we compare the dominance of the topological breakdown and the functional breakdown. During the simulation, we measure the occurrence probability $P_{topo}$ of the topological breakdown. If this probability is larger than 50\%, the topological breakdown is dominant, otherwise the functional breakdown is dominant. As a supplement to the absolute robustness studied above, the dominance is a sign of the relative robustness, i.e., which breakdown is less likely to happen than the other. The result is shown in Fig.~\ref{figure4} as a function of $\rho$. For all the network models, the topological breakdown is dominant when $\rho$ is small, and the functional breakdown is dominant when $\rho$ is close to $1$. There is a shift in the dominance, which corresponds to a particular value of $\rho$. The smaller this value is, the better relative topological robustness a network model has. As the figure shows, the NW model has the best relative topological robustness as well as the best absolute topological robustness, and the BA model has the best relative functional robustness as well as the best absolute functional robustness. Though the ER model has both the worst absolute robustnesses, its relative robustness is intermediate.

\section{the attack strategy}
In this section, edges are removed in the descending order of their betweennesses. For the sake of clarity, we first show the results for the ER model and the BA model, and then for the NW model. We study the same quantities as in the previous section.

\begin{figure}
\centering
\includegraphics[width=86mm]{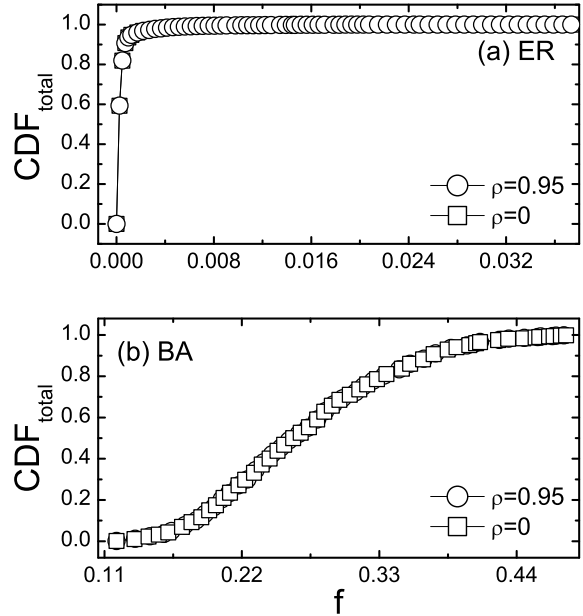}
\caption{The cumulative distribution function $CDF_{total}$ of the total breakdown under the attack strategy as a function of the fraction $f$ of removed edges for (a) the ER model and (b) the BA model. All the data is averaged over $10240$ realizations. The plots corresponding to empty load and high load almost overlap for both of the network models.} \label{figure5}
\end{figure}

The cumulative distribution function $CDF_{total}$ of the total breakdown was plotted as a function of $f$ in Fig.~\ref{figure5} for the ER model and the BA model. We first discuss the topological robustness, which corresponds to $\rho=0$. The ER networks are very vulnerable to attack, the removal of only $0.5\%$ edges almost surely destroys the total connectedness. The vulnerability is due to the lack of loops. With nearly no loops~\cite{ref1}, the ER networks can be roughly seen as trees, i.e., there is only one self-avoiding path connecting a pair of vertices. The only path is at the same time the shortest path. Many shortest paths concentrate on the high betweenness edges, which are removed by the attacker with high priority. The loss of the edges prevents communication between vertex pairs easily, as there are no alternative paths. On the other hand, the BA networks are much more robust, they can afford the removal of nearly half of the edges before almost surely break down. There is a strong correlation~\cite{ref8} in BA networks that $C_B(e)\sim k_e$, where $C_B(e)$ is the betweenness of an edge $e$ and $k_e$ is the product of the degrees of the two vertices connected by the edge. This correlation implies that at the early stage of the attack, when the characteristics of the networks are not affected too much, only the edges between vertices with large degree are targeted. The removal of these edges generally does not destroy the total connectedness and the BA networks can be reduced up to tree-like. The topological robustness of the ER model and the BA model has been studied by examining the size of the largest component in Ref.~\cite{ref8}. The authors focused on the effects of various attack strategies. Here we examine the total connectedness and provide explanations for the effects of the specific strategy. Moreover, the topological robustness is compared with the functional robustness. As shown in the figure, plots corresponding to empty load and high load almost overlap for both of the network models, indicating that the functional breakdown hardly occurs. We then investigate the average network capacity.

\begin{figure}
\centering
\includegraphics[width=86mm]{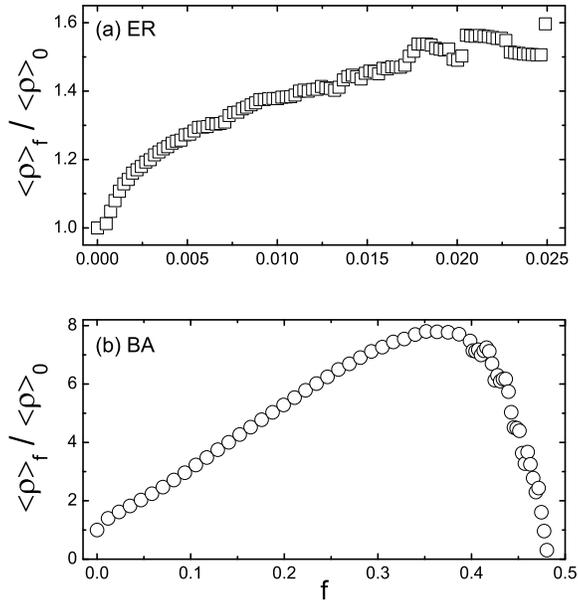}
\caption{The normalized average capacity $\left\langle \rho _c\right\rangle _f / \left\langle \rho _c\right\rangle _0$ under the attack strategy as a function of the fraction $f$ of removed edges for (a) the ER model and (b) the BA model. All the data is averaged over $10240$ realizations.} \label{figure6}
\end{figure}

The normalized average capacity $\left\langle \rho _c\right\rangle _f / \left\langle \rho _c\right\rangle _0$ was plotted as a function of $f$ in Fig.~\ref{figure6} for the ER model and the BA model. When the edge with the highest betweenness is attacked, the shortest paths that originally pass through the edge take detours. The highest edge betweenness is dispersed and the network capacity is increased. This is the case for both of the network models. The capacity of the ER networks monotonically increases during the whole process, indicating that the networks are free of the functional breakdown. The capacity of the BA networks keeps increasing until a certain value of $f$, after which the shortest paths revert to collect on the remaining edges and the capacity is decreased. As shown in the figure, the capacity is boosted up to nearly 8 times as the initial value at $f=0.35$ and then drops. We observed in the simulation that only in few realizations does the capacity drop below the initial value, thus the BA networks are almost free of the functional breakdown. Note that the growth and decay of the capacity was studied in Ref.~\cite{ref11} with the purpose of finding the optimum transmission efficiency. At each time step, the author chose to remove the edge with the largest weight, which is the product of the betweennesses of the two vertices an edge connects, and intentionally avoided the disintegration of network. Though there are some variations in the models, our result is qualitatively no different.

An interesting point worthy of note is that the attack strategy does not fulfill its name for the BA model. By comparing Fig.~\ref{figure5}(b) with Fig.~\ref{figure1}(b), we find that the BA model is more robust under the attack strategy than under the random failure strategy for any value of $\rho$. This result coincides with the moderated vulnerability of networks in terms of edge~\cite{ref13} and is in contrast to the higher vulnerability of the BA model under vertex attack than error when the existence of the giant component is used as the criterion of breakdown~\cite{ref5}.

\begin{figure}
\centering
\includegraphics[width=86mm]{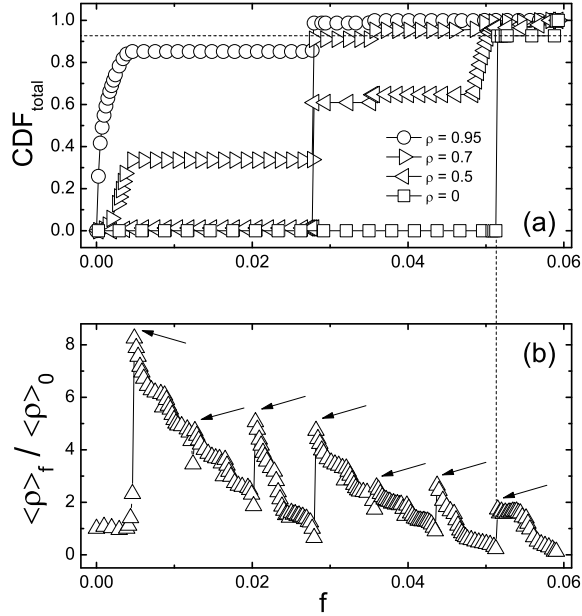}
\caption{The simulation results of the attack strategy for the NW model: (a) The cumulative distribution function $CDF_{total}$ of the total breakdown as a function of the fraction $f$ of removed edges for different values of the normalized generation rate $\rho$; (b) The normalized average capacity $\left\langle \rho _c\right\rangle _f / \left\langle \rho _c\right\rangle _0$ as a function of $f$. All the data is averaged over $10240$ realizations. The arrows point out the seven peaks of the network capacity. The dashed lines are guides to the eye: the topological breakdown ($\rho =0$) takes place just before the seventh peak for about $92.7\% $ model realizations.} \label{figure7}
\end{figure}

\begin{figure}
\centering
\includegraphics[width=86mm]{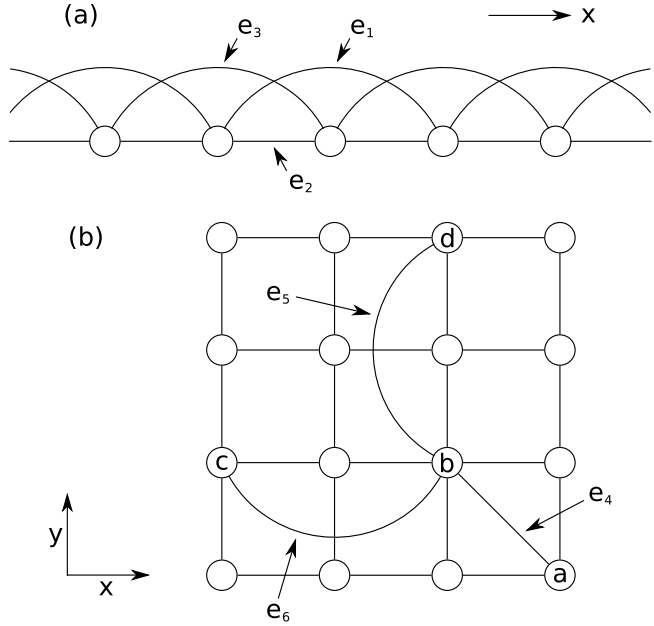}
\caption{A NW network after all the shortcuts are removed, which is a two dimensional periodic lattice with coordination number $8$, is put on a $x$-$y$ plane. Both the arcs and lines represent edges, and the circles represent vertices. (a) A slice of the lattice in the $x$ direction is shown. Every edge in the complete lattice has equal betweenness. If $e_1$ is removed, the shortest paths going from the left-hand side to the right-hand side have fewer choices of edge than before. Thus $e_2$ and $e_3$ collect larger betweenness than the others and are the immediate targets in the following attack. Note that $e_1$, $e_2$ and $e_3$ each represent a group of $L$ edges in the $y$ direction. (b) The edge $e_4$ is such a special shortcut that it spans less lattice distance than the next-neighbor links and collects fewer shortest paths. Though the betweenness of the shortcut is low, the betweennesses of the neighboring edges are increased, e.g., the shortest paths connecting $a$, $c$ and $a$, $d$ both pass through $b$, the edges $e_5$ and $e_6$ are biased. Different from $e_6$, the edge $e_5$ is in the $y$ direction. In addition to the six peaks in the $x$ direction, one more peak of the network capacity could be seen. In fact the effect of the shortcut $e_4$ is weak that the counterpart of $e_5$ which connects to $a$ in the $y$ direction is not removed before the lattice is split. For clarity, some next-neighbor links are omitted.} \label{figure8}
\end{figure}

The cumulative distribution function $CDF_{total}$ of the total breakdown and the normalized average capacity $\left\langle \rho _c\right\rangle _f / \left\langle \rho _c\right\rangle _0$ are shown as functions of $f$ for the NW model in Fig.~\ref{figure7}. The figure can be interpreted as follows: Starting from a complete NW network, the attacker first targets the shortcuts, because they collect the most shortest paths. After all the shortcuts are removed, the rest of the network is a two dimensional $L\times L$ periodic lattice with each vertex connected to its neighbors and next neighbors. The structure becomes completely homogeneous and the network capacity is boosted. Thus we see the first peak of the network capacity at $f=f_1=2p/\langle k\rangle =0.005$ in Fig.~\ref{figure7}(b). The attacker continues by randomly targeting an edge. We show a diagram in Fig.~\ref{figure8}(a) to help the interpretation. The lattice is put on a $x$-$y$ plane and Fig.~\ref{figure8}(a) shows a slice along the $x$ direction. For every edge in the figure, there are $L-1$ parallel counterparts in the $y$ direction. The $L$ edges form a group that if any edge in the group is removed, the others are removed in the next few steps as well, because shortest paths that originally pass through the whole group concentrate on the remaining members, raising the betweennesses of them. Suppose that edge $e_1$ in the figure is removed, the network capacity drops until all the counterparts of the edge are removed. The heterogeneity of the raised betweenness is alleviated and we see the second peak of the network capacity, which is $L$ edges away from the first peak. However, the capacity is not fully restored, as edge $e_2$, $e_3$ and their counterparts collect larger number of shortest paths than before. Thus two peaks follow, each at an interval of $L$ edges. With the $3L$ edges gone, the lattice is no longer periodic and the center area in the $x$ direction collects the highest betweenness. Then the following attack removes another $3L$ edges in this area and the network is split. For the most of the model realizations, six peaks of the network capacity are observed, which conforms to the arguments above. There is a special case that allows the observation of one more peak. As depicted in Fig.~\ref{figure8}(b), it is possible that a shortcut $e_4$ is placed between vertex $a$ and $b$. This kind of shortcut spans less lattice distance than the abundant next-neighbor links, thus collects less shortest paths and is removed very late in the attack. The existence of such shortcut brings higher betweenness to a group of $L$ edges in the $y$ direction, e.g., all the shortest paths passing through $a$ and $c$ collect on the edge $e_6$ in the figure. Removing this group of edges induces the seventh peak. The seven peaks are pointed out in Fig.~\ref{figure7}(b) by arrows. We see in the figure that the network capacities on the peaks are larger than the initial value, thus only when the capacities in the valleys drop below the initial value does the network functionally break down, i.e., $CDF_{total}$ shown in Fig.~\ref{figure7}(a) resembles stairs for $\rho \neq 0$. When $\rho =0.95$, nearly $90\%$ model realizations functionally break down before $f=0.005$. The situation is only a little better than the ER model and much worse than the BA model. For the topological case $\rho =0$, there is a plateau in the figure which corresponds to the emergence of the seventh peak. We estimate the emergence probability as ${pN \choose 1}P_{sc}\left(1-P_{sc}\right)^{p N-1}\approx 7.3\%$, where $P_{sc}=\frac{4}{N-8}$ is the probability of the occurrence of the special shortcut and we only consider the first order of this probability. The result is supported by Fig.~\ref{figure7} that the topological breakdown takes place just before the seventh peak of the network capacity, at $f=f_7=f_1+6L/(\langle k\rangle N/2)=0.0519$, for about $92.7\% $ model realizations. The topological breakdown for the rest of the model realizations takes place after $L$ more edges are removed. The NW model has intermediate topological robustness between the ER model and the BA model.

\begin{figure}
\centering
\includegraphics[width=86mm]{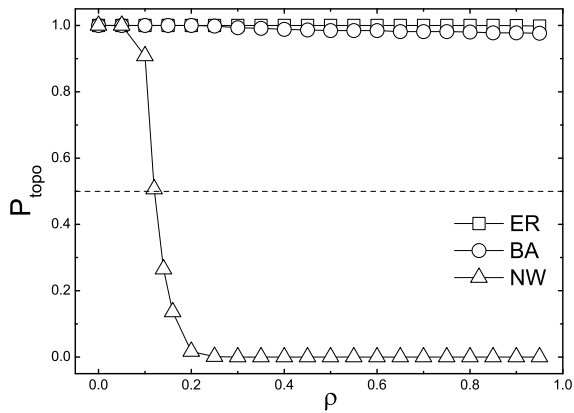}
\caption{The occurrence probability $P_{topo}$ of the topological breakdown under the attack strategy as a function of the normalized generation rate $\rho$ for the network models. The data is averaged over $10240$ realizations. The dashed line is for $P_{topo}=50\%$. The network model that corresponds to the smallest $\rho$ at the point of intersection between $P_{topo}$ and this line has the best relative topological robustness.} \label{figure9}
\end{figure}

The relative robustness of the network models for the attack strategy is studied in the same way as in the previous section for the random failure strategy. The result is shown in Fig.~\ref{figure9}. The ER model has the worst absolute topological robustness, and also the worst relative topological robustness, as for any value of $\rho$, no functional breakdown takes place. While the BA model has the best absolute topological robustness, the relative topological robustness is still bad, nearly the same as the ER model. For the NW model, we know that the topological breakdown only takes place when at least $6L$ edges are removed, but the functional breakdown could happen if network capacity drops below the initial value. The figure shows that for $\rho >0.2$, the topological breakdown hardly takes place. Thus the NW model has the best relative topological robustness.

\section{conclusion}
In this paper, we have studied the topological robustness and the functional robustness of several network models under two strategies of edge removal, using the total connectedness as the measure. For each removing strategy, we have examined the combined effect and the relative importance of the two types of robustness. Through the study of the combined effect, we have found out the network topology which is the most robust in a specific environment, e.g., the NW model is the most robust in the environment of the random edge removal, while the BA model is the most robust in the environment of the edge attack. Through the study of the relative importance, we have known with evidence how to strengthen a network efficiently, e.g., as the NW model has the best relative topological robustness, improvements on the functional robustness should be emphasized, such as increasing the capacity of vertices; the BA model has bad relative topological robustness under the edge attack, thus enhancements on the connectivity such as building redundant links are appropriate. These results can have applications in designing and optimizing artificial networks, such as implementing a robust P2P network where a connection between peers has a constant probability to fail. There are also some extensions for further studies. The traffic dynamics in our model is far from realistic, more elements abstracted from real-world traffic can be incorporated. Moreover, networks are not limited to carrying traffic. It is an open question how the interaction between the topologies and different dynamics influences the network robustness.

\begin{acknowledgements}
 The work was supported by the National Natural Science
Foundation of China under Grant No. 10334020.
 \end{acknowledgements}

\end{document}